# Organic Photodiodes with an Extended Responsivity using Ultrastrong Light-Matter Coupling

*Elad Eizner*[*], *Julien Brodeur, Fábio Barachati, Aravindan Sridharan, Stéphane Kéna-Cohen*[*]

Department of Engineering Physics, École Polytechnique de Montréal, Montréal H3C 3A7, QC, Canada.

**ABSTRACT**

In organic photodiodes (OPDs) light is absorbed by excitons, which dissociate to generate photocurrent. Here, we demonstrate a novel type of OPD in which light is absorbed by polaritons, hybrid light-matter states. We demonstrate polariton OPDs operating in the ultra-strong coupling regime at visible and infrared wavelengths. These devices can be engineered to show narrow responsivity with a very weak angle-dependence. More importantly, they can be tuned to operate in a spectral range outside that of the bare exciton absorption. Remarkably, we show that the responsivity of a polariton OPD can be pushed to near infrared wavelengths, where few organic absorbers are available, with external quantum efficiencies exceeding those of a control OPD.

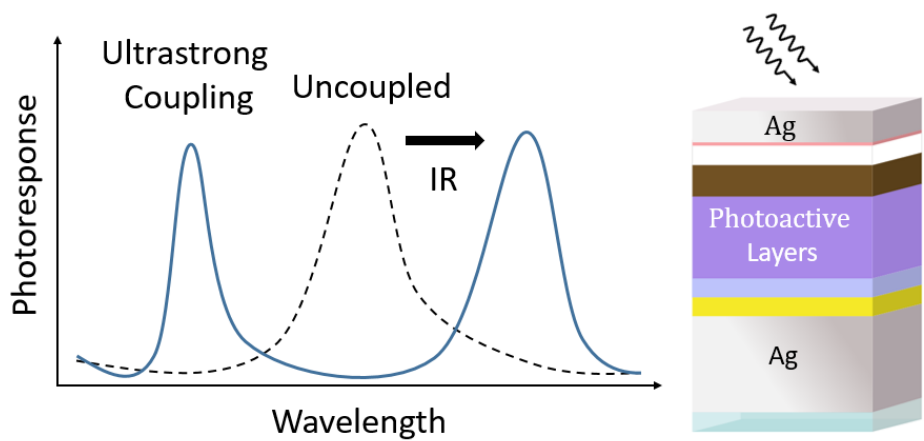

**Introduction.**

The quantum vacuum and its fluctuations are at the heart of some of the most important effects in physics such as spontaneous emission, van der Waals and Casimir forces, and the Lamb shift[1,2]. Recent results indicate that the vacuum electromagnetic (EM) field can also be used to dramatically modify the optical and electrical properties of molecules[3–6]. The ability to modify molecular properties is fascinating not only from a fundamental standpoint, but also directly relevant to improving organic optoelectronic devices.

These modifications occur when molecules are placed within an optical cavity such that their electronic transition can interact strongly with the vacuum field. In this regime of interaction, the rate of energy transfer between Frenkel molecular excitons and EM modes is faster than their individual uncoupled decay rates and hybrid light-matter quasiparticles termed polaritons are formed. Over the last decade, organic polaritons have attracted great interest due to their remarkable light-matter properties, which were shown to lead to fascinating physical phenomena at room temperature, e.g. Bose–Einstein condensation[7–9] and superfluidity[10]. In addition, novel device applications are emerging such as low-power polariton lasers, polariton optical circuits and quantum devices[5].

The polariton dispersion relation is separated into upper (UP) and lower (LP) branches with resonance energies that lie on opposite sides of the bare exciton energy. The minimum splitting between the LP and UP energies is known as the vacuum Rabi splitting energy ($\hbar\Omega$) and quantifies the light-matter interaction strength. This energy $\hbar\Omega \propto f\sqrt{N/V}$, where $f$ is the oscillator strength of the excitonic transition and $N$ is the number of molecules that are within the EM mode volume $V$. When the Rabi splitting becomes a significant fraction of the uncoupled exciton energy ($\hbar\Omega >$ $0.2E_x$), the system is in the ultra-strong coupling (USC) regime[11–14]. In this regime, the Jaynes-Cumming Hamiltonian breaks down (i.e. dipolar interaction with a rotating wave approximation) and additional interaction terms due to anti-resonant coupling and the contribution from the square magnetic field vector must be taken into account[12] . Some of the remarkable new features that arise from the full Hopfield Hamiltonian treatment are an asymmetric anti-crossing as a function of $\Omega$ and a ground state with finite photon and exciton content. From a device perspective, USC also leads to a spectral response that is nearly angle-independent despite its sharp resonant behavior.

In organic photodiodes, excitons are first generated by light absorption in an organic active layer. They then diffuse to a donor-acceptor interface where they can dissociate into free carriers. Finally, the charges are transported to opposing electrodes of the device due to the internal electric field of the structure. The efficiency and spectral response of an OPD is of course limited by the photophysical and optoelectronic properties of the active layers. In particular, it has been extremely challenging to develop active layers (typically the donor) which absorb light efficiently at near infrared and infrared wavelengths. Based on the previous discussion, an interesting approach would then be to hybridize the donor exciton with the vacuum EM field in order to further push its absorption to the infrared without requiring the synthesis of new materials. In the USC regime, the LP absorption can be red-shifted by a significant fraction of the bare exciton energy while still maintaining a large exciton content.

In this paper, we demonstrate polariton OPDs operating in the USC regime. This is achieved in a standard OPD structure consisting of a mixed donor-acceptor heterojunction with Silver (Ag) electrodes that also act as mirrors to form the optical cavity. We study two types of polaritonic OPDs using donor materials that have previously been widely used for organic photovoltaics. A visible absorber, copper (II) phthalocyanine (CuPc), and a near infrared absorber, tin (II) phthalocyanine (SnPc). Remarkably, we achieve large Rabi splittings of up to 0.7 eV (42% of the exciton transition energy) and show that the LP can be pushed to wavelengths beyond 1 μm with external quantum efficiencies (EQEs) exceeding those of the control device. Furthermore, given that polaritons are resonantly excited in these devices, they are ideal platforms for studying potential modifications in diffusion lengths, carrier mobility and charge separation in the strong light-matter coupling regime.

## Results and Discussion.

Figure 1a illustrates schematically how in the USC regime the uncoupled absorption of the photoactive layers in an OPD (dashed line) can be separated into LP and UP absorption bands to address new spectral ranges (solid line). The planar-mixed polariton photodiode architecture studied in this work is shown in Figure 1b.

**Visible Polariton Photodiodes.**

To realize polariton OPDs that operate in the visible region of the spectrum, we chose CuPc as the donor. Copper (II) phthalocyanine is a bright blue synthetic pigment that has been an archetypal donor for organic photovoltaics, where the combination of CuPc and $C_{60}$ layers is often considered as a model system to study molecular donor-acceptor processes[17–21]. The normalized absorbance of a CuPc thin film is shown in Figure 1c as a dashed blue line. It exhibits a double peaked absorbance with a maximum at $\lambda = 628$ nm. The normalized absorbance for a mixed film of CuPc and $C_{60}$ (70%: 30% by volume) is shown in Figure 1c as a solid blue line. The ratio between the two peak values in the mixed film is reduced, but the spectral positions of the peaks are only slightly modified. The polariton OPD structure consists of a Ag anode (75 nm), 1,4,5,8,9,11-hexaazatriphenylene hexacarbonitrile (HATCN; 15 nm), CuPc (15 nm), Mixed CuPc:$C_{60}$ (60 nm, 73%: 27% by volume), $C_{60}$ (25 nm), 4,7-diphenyl-1,10-phenanthroline (BPhen; 15 nm), lithium fluoride (LiF; 1 nm) and an Ag cathode (25 nm).

To extract the polariton dispersion relation, we measured angle-resolved reflectivity ($R$) on the fabricated photodiode with light incident from the cathode direction. Given the use of a thick 75 nm bottom Ag electrode, transmission can be neglected and the absorption can be calculated as $A = 1 - R - T \approx 1 - R$. The angle-resolved absorption is shown for transverse electric (TE) and transverse magnetic (TM) light polarization in Figures 2a, 2b, respectively. For each incident angle we observe two peaks corresponding to the formation of LP and UP states in the USC regime. The black lines show the least square fit of the peak values to the Hopfield Hamiltonian (see Methods). The dashed blue lines show the uncoupled exciton energy ($E_x$) and the cavity photon energies, where the cavity energy at normal incidence is $E_c$. From the fit, we obtain a Rabi splitting $\hbar\Omega = 0.55 \pm 0.01$ eV with detuning of $\Delta = E_c - E_x = -0.18 \pm 0.02$ eV for the TE mode and $\hbar\Omega = 0.52 \pm 0.01$ eV with detuning of $\Delta = -0.20 \pm 0.02$ eV for the TM mode. The Rabi energy corresponds to 28% of the uncoupled exciton energy ($\hbar\Omega/E_x$). To compare our results to those of a passive microcavity, we also fabricated a microcavity consisting of Ag (75 nm), CuPc (150 nm) and Ag (35 nm) on a glass substrate. The angle resolved reflectivity measurements for TE and TM polarizations are shown in Figure S1. In this case, the Rabi splitting reaches a value of $\hbar\Omega = 0.95 \pm 0.01$ eV which is 48% of the uncoupled exciton energy (see Supplementary Information).

To study the photodiode performance, we measured the angle-resolved photocurrent spectra of the device and calculated the corresponding EQE (see Methods). Figure 2c shows the angle-resolved absorption for unpolarized light and the angle-resolved EQE is shown in Figure 2d. As can be seen, there is a one-to-one correspondence between the dispersion of the UP and LP modes measured in absorption and the peaks in the EQE spectra.

To compare the results to a photodiode in the weak coupling regime, an identical structure was prepared using an indium tin-oxide (ITO) anode instead of the 75 nm Ag layer. Figure 3a shows the EQE spectra for the control device with light incident from the ITO-side at normal incidence (dashed black line), and for the polaritonic OPD at normal incidence (solid blue line) and at 60 degrees (solid red line). We observe that the polariton EQEs do not surpass those of the control device given the significantly higher absorption in the active layers for light incident from the ITO-side compared to the case where light is incident through the cathode. Nevertheless as we will show in the next section that it is possible to exceed the EQE of our control devices by adequately choosing the detuning between the exciton and photon resonances.

We further analyze the performance by calculating the internal quantum efficiencies (IQEs) of the devices. This serves to remove the effect of light absorption from the calculated efficiency. Figure 3(b) show the polariton IQEs at wavelengths corresponding to the LP and UP resonances, and between the modes at $\lambda = 660$ nm, where uncoupled excitons (partially dark states) are principally excited. As can be seen, the polariton IQEs are lower, but comparable to the ones at $\lambda = 660$ nm and to those of the control device at $\lambda = 630$ nm. Also, we note that the IQE of the UP decrease as a function of the incident angle. The qualitative agreement in IQE for the polaritonic modes, the control and mid-gap excitation suggests that polaritons first scatter into the dark state manifold prior to dissociation.

Small changes in IQE can be understood based on the spatial distribution of the absorbed photons within the device. To study these effects, we have calculated the dissipated power of the incident EM field within the device (see Methods). Figure 3c shows the distribution of the EM dissipated power within the different layers that compose the polariton OPD. As can be seen, power is dissipated not only in the active layers, but also in the Ag electrodes as Joule heating. Moreover, the distribution of the dissipated power in the CuPc, $C_{60}$ and mixed layers is very different for UP and LP modes. In particular, the UP power is mostly dissipated in the planar CuPc and $C_{60}$ layers, which leads to inefficient photocurrent generation. Figure 3d show the ratio between the power dissipation in the mixed CuPc:$C_{60}$ and the total dissipated power in the device. The angular dependence of this ratio shows a qualitative behavior similar to the IQEs for the LP and UP. Also, the relative power dissipation is higher for the control device at $\lambda = 630$ nm.

**Infrared Polariton Photodiodes.**

To demonstrate that the LP can be pushed to infrared wavelengths where there is currently a lack of efficient organic absorbers we use SnPc[22–24] as the donor material. The normalized absorbance of a SnPc thin film is shown in Figure 1c as a dashed black line. It exhibits two peaks at $\lambda = 745$ nm and $\lambda = 885$ nm, previous works have shown that these peaks originate from monomers and aggregates absorption respectively[22–24]. The normalized absorbance for a mixed film of SnPc and $C_{60}$ (70%: 30% by volume) is shown in Figure 1c as a solid black line. Notably, when mixed with $C_{60}$, the monomer absorption dominates.

We fabricated two infrared polariton OPDs by varying the thickness of the mixed layer. The device structure consisted of Ag (75 nm), HATCN (15 nm), SnPc (10 nm), mixed SnPc:$C_{60}$ (75 nm or 110 nm; 73%: 27% by volume), $C_{60}$(30 nm), BPhen (15 nm), LiF (1 nm) and Ag (25 nm). The angle-resolved absorption for the 75 nm-thick mixed layer is shown in Figure 4a. The black lines show the least square fit of the TM polarization peak values to the Hopfield Hamiltonian model. A Rabi splitting of $\hbar\Omega = 0.58 \pm 0.01$ eV and a detuning of $\Delta = -0.13 \pm 0.02$ eV are found. The Rabi splitting corresponds to 35% of the uncoupled monomer exciton energy. The angle-resolved EQE spectra of the device are shown in Figure 4b. Note that as previously reported, SnPc OPDs show much lower EQEs than their CuPc counterparts due to a much lower hole mobility[23].

The angle-resolved absorption for the 110 nm-thick mixed layer is shown in Figure 4c. We find a Rabi splitting of $\hbar\Omega = 0.70 \pm 0.01\ eV$ and a detuning of $\Delta = -0.25 \pm 0.01$ eV. The Rabi splitting correspond to a notable 42% of the uncoupled monomer exciton energy. The angle-resolved EQE of the device is shown in Figure 4d. Remarkably, the LP operation extends to beyond 1 μm with an EQE much higher than that of the control device at the same wavelengths. We also compared the Rabi energy to a passive microcavity consisting of Ag (75 nm), mixed SnPc:$C_{60}$(160 nm; 70%: 30% by volume) and Ag (20 nm) on a glass substrate. In this case, the angle resolved reflectivity for TE and TM polarizations are shown in Figure S2. The found Rabi splitting is $\hbar\Omega = 0.78 \pm 0.01$ eV which correspond to 47% of the uncoupled monomer exciton energy (see Supplementary Information).

Figure 5a shows the EQE of the 75 nm device at normal incidence and at 60 degrees (blue and red lines), and of the control device (dashed black line). The corresponding IQEs of the polariton OPD at the LP and UP wavelengths, and within the gap at $\lambda = 750$ nm are shown in Figure 5b. The IQEs of the UP have similar values as the ones at λ = 750 nm and are comparable to the IQE of the control device at the EQE peak ($\lambda = 720$ nm). The IQEs of the LP and also of the aggregate peak in the control device ($\lambda = 880$ nm) are considerably higher. Simulations of the dissipated power in the polariton OPD show that for incident light at an angle of 40 degrees, 41% of the light is dissipated in the 10 nm planar SnPc layer, while only 8% for the UP. In addition, 11.5% of the light power will be dissipated in the Ag electrodes for the LP resonance compared to 6% for the UP. Hence, we see that for wavelengths that overlap with the absorption of SnPc aggregates, there is a large contribution to the EQE and IQE from aggregates in the bare film.

Figures 5c, 5d show the EQEs and IQEs for the infrared polariton OPD with a 110 nm-thick mixed layer. Remarkably, the EQE of the LP mode is greatly enhanced compared to the EQE of the control device at these wavelengths due to the significantly higher absorption. This demonstrates that polariton OPDs can operate efficiently at wavelengths beyond those achievable by the bare material. Also here, the IQEs of the UP have similar values as the ones at $\lambda = 750$ nm and are slightly higher than the IQE value of the control at $\lambda = 720$ nm. On the other hand, the IQEs of the LP are lower. From simulations of the dissipated power of the incident light in the device we find that for the LP, a very large portion of the light is dissipated in the Ag electrodes which explains the reduced IQE values that are observed. For an incident angle of 40 degrees, 33% of the light is dissipated in the Ag electrodes for the LP resonance, while only 6.6% for the UP. In addition, 15.5% of the light is dissipated in the 10 nm layer of SnPc for the LP resonance ($\lambda = 984$ nm), and 5% for the UP. Note that these metal losses are not inherent to polaritonic devices and can be substantially reduced by further optimization.

**Concluding Remarks.**

We developed a novel type of OPDs in which the basic excitations are not excitons but ultra-strongly coupled hybrid photon-exciton states. Given that such polaritons are resonantly excited in these devices, the structure is an ideal testing ground for modifications in molecular photophysics beyond the absorption spectrum. However, the behavior we observe can be well understood semi-classically and nothing indicates significant changes in diffusion or charge transport. Nevertheless a recent theoretical proposal has highlighted how polaritons could possibly influence singlet fission in solar cells[6]. Unlike polaritonic OLEDs, where light is emitted only from the LP mode[25], here photocurrent with maximal EQEs is observed for both the LP and the UP modes. Another beneficial feature is the low angular dependence of the devices. Finally, we show that the polariton OPD operating wavelengths can be pushed far away from the uncoupled absorbance. For SnPc, we obtain a moderate responsivity beyond 1 μm with EQEs that are much higher than that of the control device. Polariton OPDs can be further optimized and our results open the door for novel organic optoelectronic devices such as photodetectors, upconvertors and solar cells based on strong light-matter coupling.

## Methods.

**Sample Preparation.** The devices were fabricated on glass substrates using thermal evaporation at a base pressure $< 10^{-7}$ Torr (EvoVac, Angstrom Engineering). Control devices were fabricated on a glass substrate coated with 125 nm of ITO. Prior to the deposition of the films, the substrates were cleaned and exposed to a UV-ozone treatment. The organic layers were deposited at a growth rates of $0.5 - 1.5$ Å/s, the 75 nm Ag electrode was deposited at 10 Å/s and the 25 nm Ag electrode was deposited at 2 Å/s. Mixed layers were grown by co-deposition at different deposition rates and the device area was $2 \times 2$ mm. The entire device structure was fabricated without breaking vacuum. CuPc and SnPc were purchased from Lumtec (Sublimed, > 99%).

**Characterization.** The refractive index and the layer thicknesses were obtained using ellipsometry (J. A. Woollam Co., RC2 D+NIR). Angle-resolved reflectivity was performed using the same instrument with focusing probes. Angle resolved-EQE measurements were performed using a Xenon lamp light source (Oriel Instruments) connected to a Monochromator (Acton Research Corporation). The devices were positioned on a rotating stage and collimated light with spot size smaller than the device area was illuminated at spectral steps of 2 nm. For each angle and wavelength, the photocurrent (at zero bias voltage) was collected using a source measure unit (Keithley 2614B). For calibration, the spectral light power was obtained using a calibrated Silicon photodiode (Hamamatsu S3204-08). The EQEs were then calculated by the ratio between the number of collected charges to the number of incident photons (EQE = $IP\hbar\omega/e$), where $I$ is the photocurrent, $P$ is the power of incident photons, $\hbar\omega$ is a photon energy and $e$ is an electron charge. The IQEs of the polariton OPDs were calculated from the experimental peaks of the EQEs and of the absorption. The IQEs of the control device were calculated using simulated absorption values.

**Theoretical Modeling - Hopfield Hamiltonian.** The experimental Rabi splitting and the detuning were extracted for TE or TM modes by performing a least-square fit of the reflectivity dips to the following linear equation [11–13]

$$\begin{pmatrix} E_{ph}(q)+2D & \Omega/2 & 2D & \Omega/2 \\ \Omega/2 & E_x & \Omega/2 & 0 \\ -2D & -\Omega/2 & -E_{ph}(q)-2D & -\Omega/2 \\ -\Omega/2 & 0 & -\Omega/2 & -E_x \end{pmatrix} \begin{pmatrix} w_{j,q} \\ x_{j,q} \\ y_{j,q} \\ z_{j,q} \end{pmatrix} = E_{j,q} \begin{pmatrix} w_{j,q} \\ x_{j,q} \\ y_{j,q} \\ z_{j,q} \end{pmatrix} \quad (1)$$

Equation 1 is a solution of Hopfield Hamiltonian and the eigenvalues correspond to the polariton energies $\pm E_{LP,q}$ and $\pm E_{UP,q}$. The in-plane wavevector is $q$, $j \in \{LP, UP\}$ and $D = \Omega^2/4E_x$. We used a photonic cavity dispersion with an effective refractive index $n_{eff}$,

$$E_{ph}(q) = \sqrt{(\frac{hcq}{n_{eff}})^2 + E_c^2}, \quad (2)$$

with $q = \frac{\omega}{c}\sin\theta$, where $\theta$ is the angle of incidence.

**Power Dissipation.** To calculate the spatial distribution of the EM power dissipation inside the devices, the electric field at a position $x$, $E(x)$, was calculated following a transfer matrix approach for either TE or TM light polarization[26]. The dissipated power is calculated using the time average of the energy dissipated per second in each layer from

$$Q(x) = \frac{\varepsilon_0 \pi c}{\lambda} \text{Im}[\varepsilon] |E(x)|^2, \quad (3)$$

where $\epsilon_0$ is the vacuum permittivity and $\epsilon$ is the dielectric constant of the corresponding layer.

**Supporting Information**. Angle resolved reflectivity spectra of ultra-strongly coupled organic microcavities containing CuPc or a mixed layer of SnPc and $C_{60}$.

AUTHOR INFORMATION

*Corresponding authors: E.E. elad.eizner@polymtl.ca, S. K-C. s.kena-cohen@polymtl.ca.

Notes

The authors declare no competing financial interest.

ACKNOWLEDGMENT

The authors acknowledge funding from the NSERC Discovery Grant Program (RGPIN-2014-06129) and the Canada Research Chairs program.

.

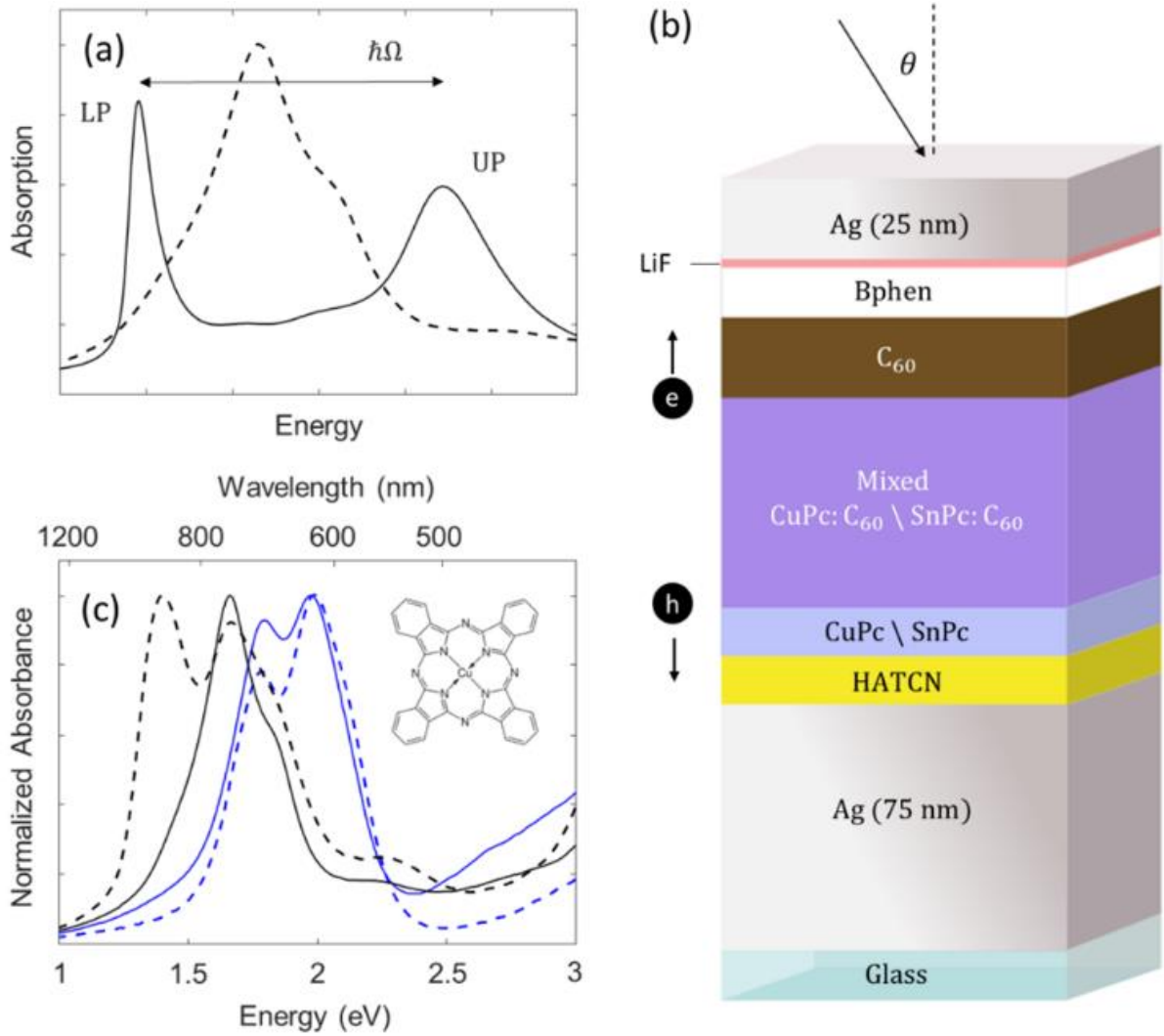

**Figure 1.** (a) Absorption spectrum of the uncoupled photoactive layers (dashed) and the ultra-strongly coupled system (solid). (b) Polariton photodiode structure. (c) Normalized absorbance spectra of 50 nm thin films on a quartz substrate. Dashed blue line: CuPc, solid blue line: mixed CuPc: $C_{60}$ (70%:30% by volume), dashed black line: SnPc and solid black line: mixed SnPc: $C_{60}$ (70%:30% by volume). The inset shows the molecular structure of CuPc.

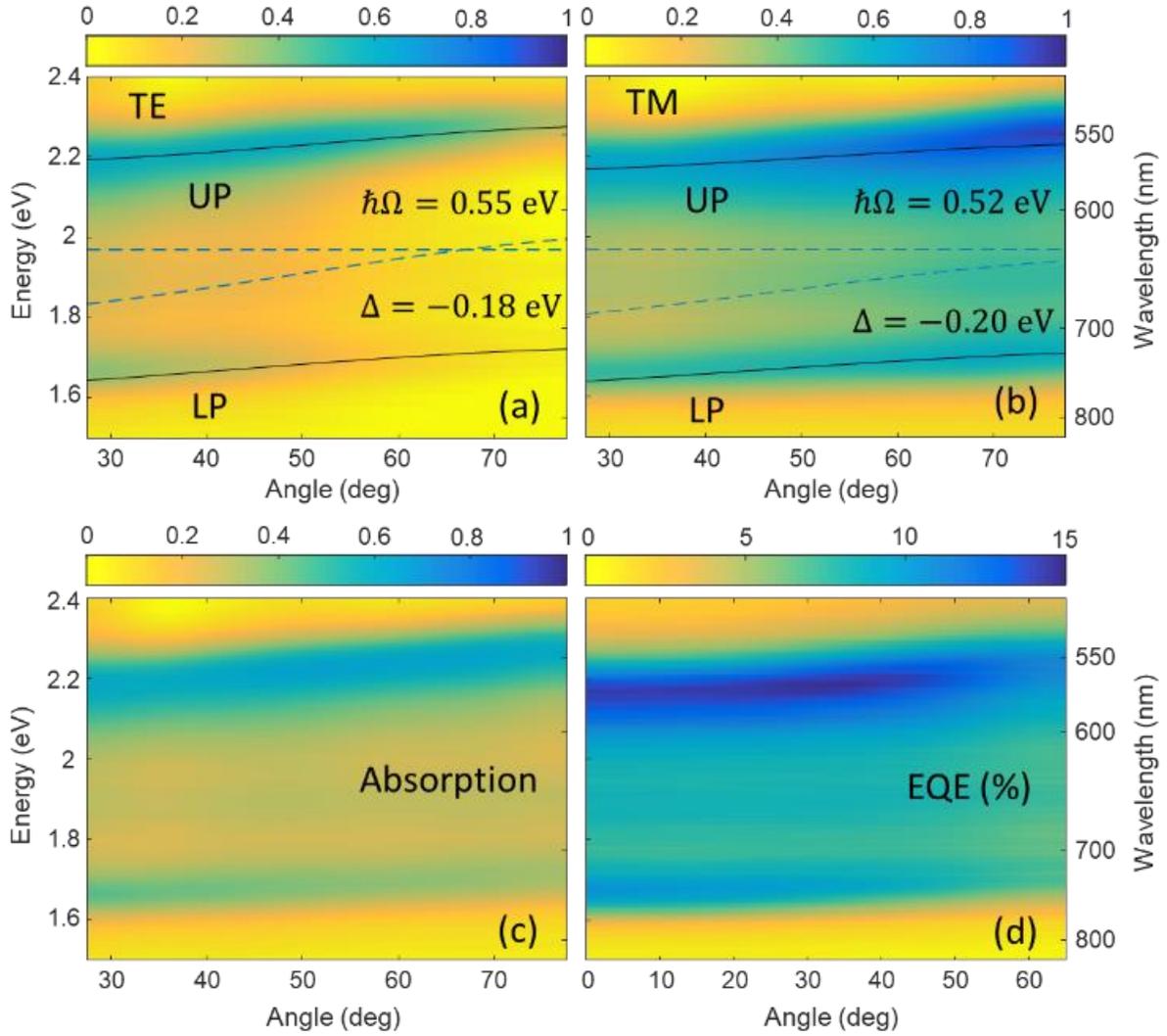

**Figure 2.** Characterization of the visible polariton photodiode. Angle resolved absorption spectra for (a) TE polarization, (b) TM polarization and (c) unpolarized light. The solid black lines show the least square fit to the Hopfield Hamiltonian. The dashed lines are the energies of the corresponding uncoupled photonic and excitonic modes. (d) Angle resolved EQE spectra.

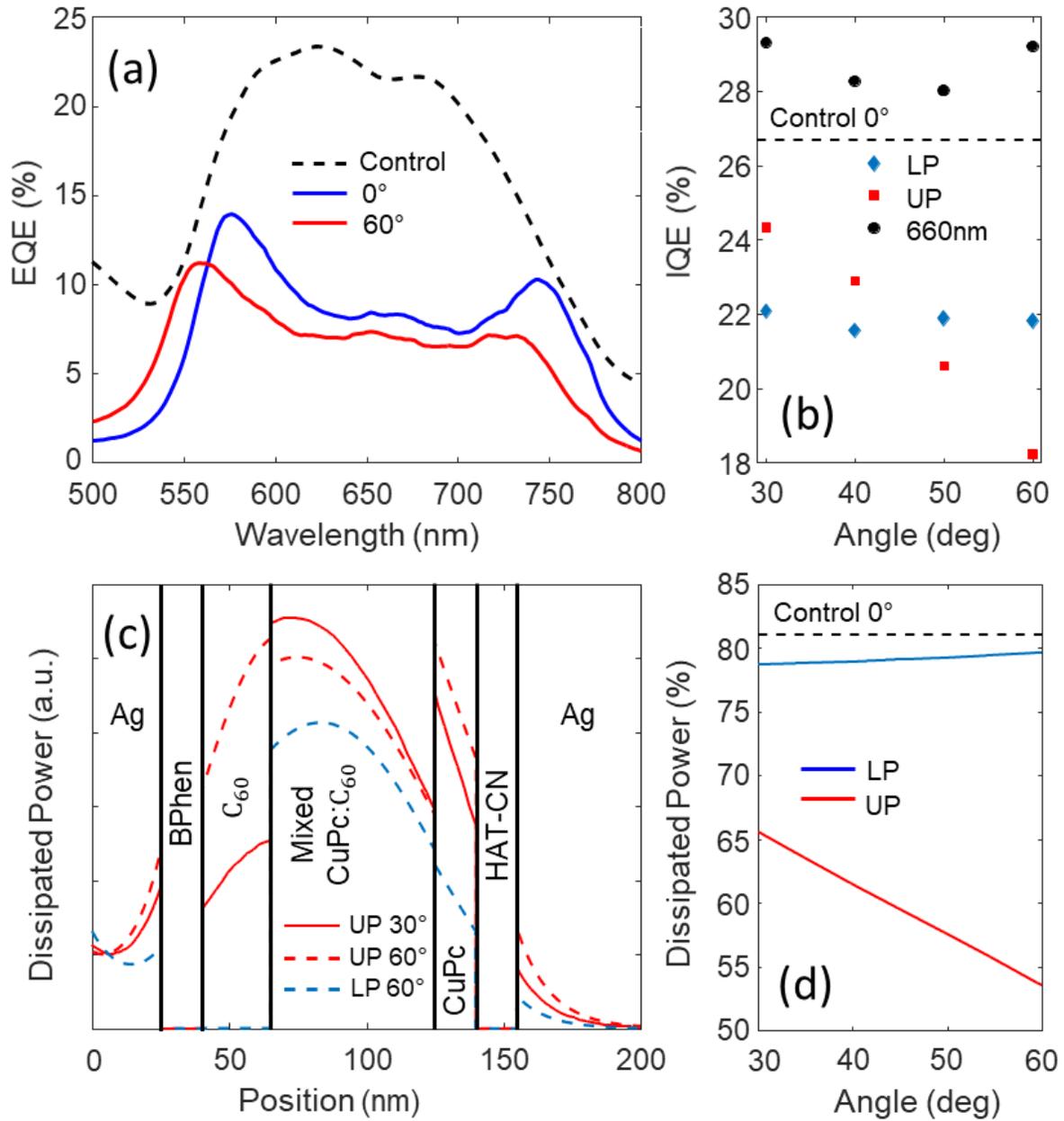

**Figure 3.** (a) EQE spectra for the visible polariton photodiode with incident light normal (blue line) and at 60 degrees (red line). Dashed black line: EQE spectrum for the control device with normal incident light shined from the ITO-side. (b) IQE vs angle. The IQE of the control device was calculated at $\lambda = 630$ nm. (c) Theoretical calculation of the power dissipated in each layer of the photodiode structure by the electromagnetic field. (d) The ratio between the dissipated power in the mixed CuPc:$C_{60}$ layer and the total dissipated power in the device.

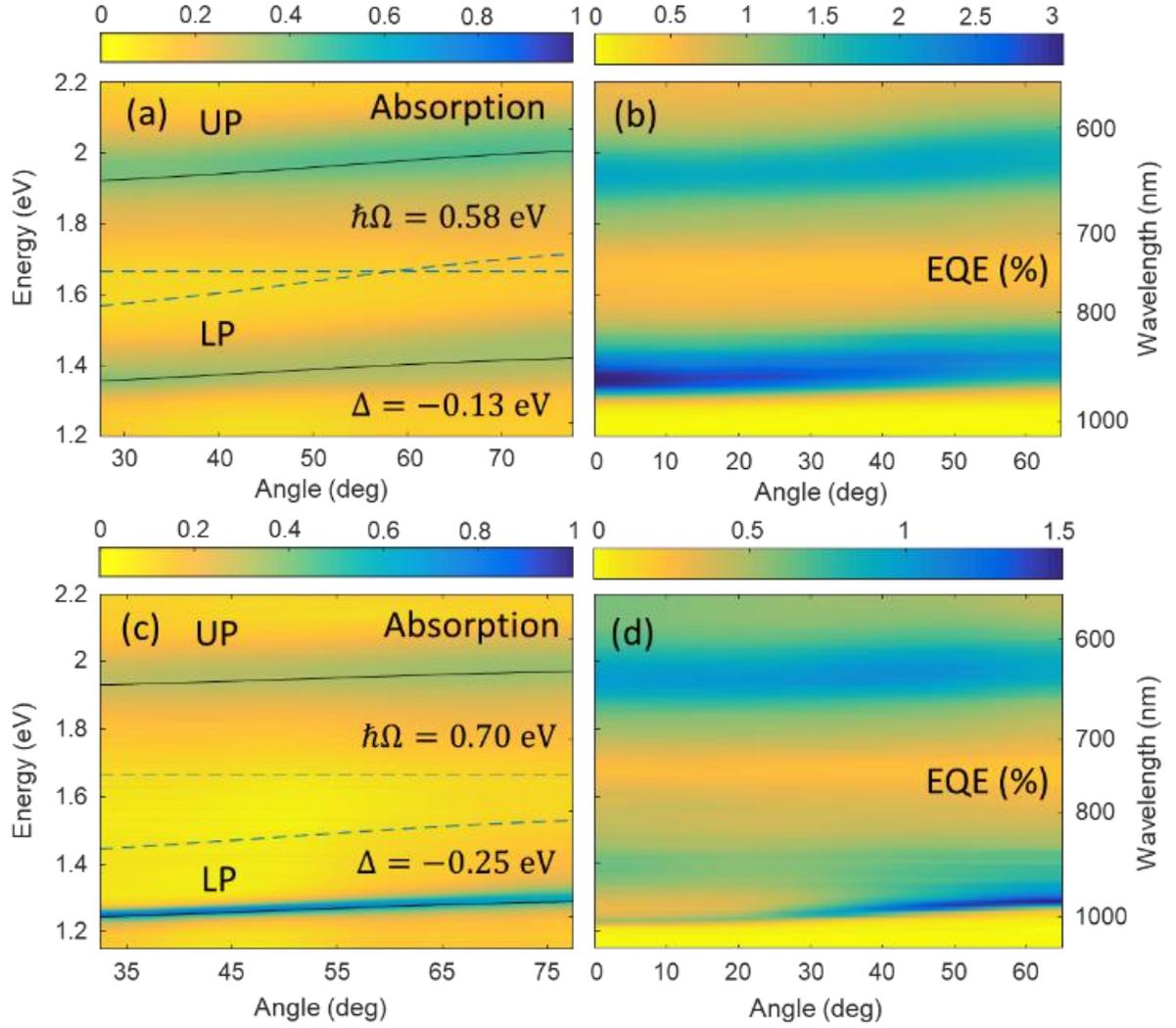

**Figure 4.** Characterization of the infrared polariton photodiodes. (a) Angle resolved absorption spectra using unpolarized light for the photodiode with mixed SnPc: $C_{60}$ layer thickness of 75 nm. The solid black lines show the least square fit of TM polarized reflectivity to the Hopfield Hamiltonian. The dashed lines are the energies of the corresponding uncoupled photonic and excitonic modes. (b) Angle resolved EQE spectra of the device. (c) Angle resolved absorption spectra for the photodiode with mixed SnPc: $C_{60}$ layer thickness of 110 nm. (d) Angle resolved EQE spectra of the device.

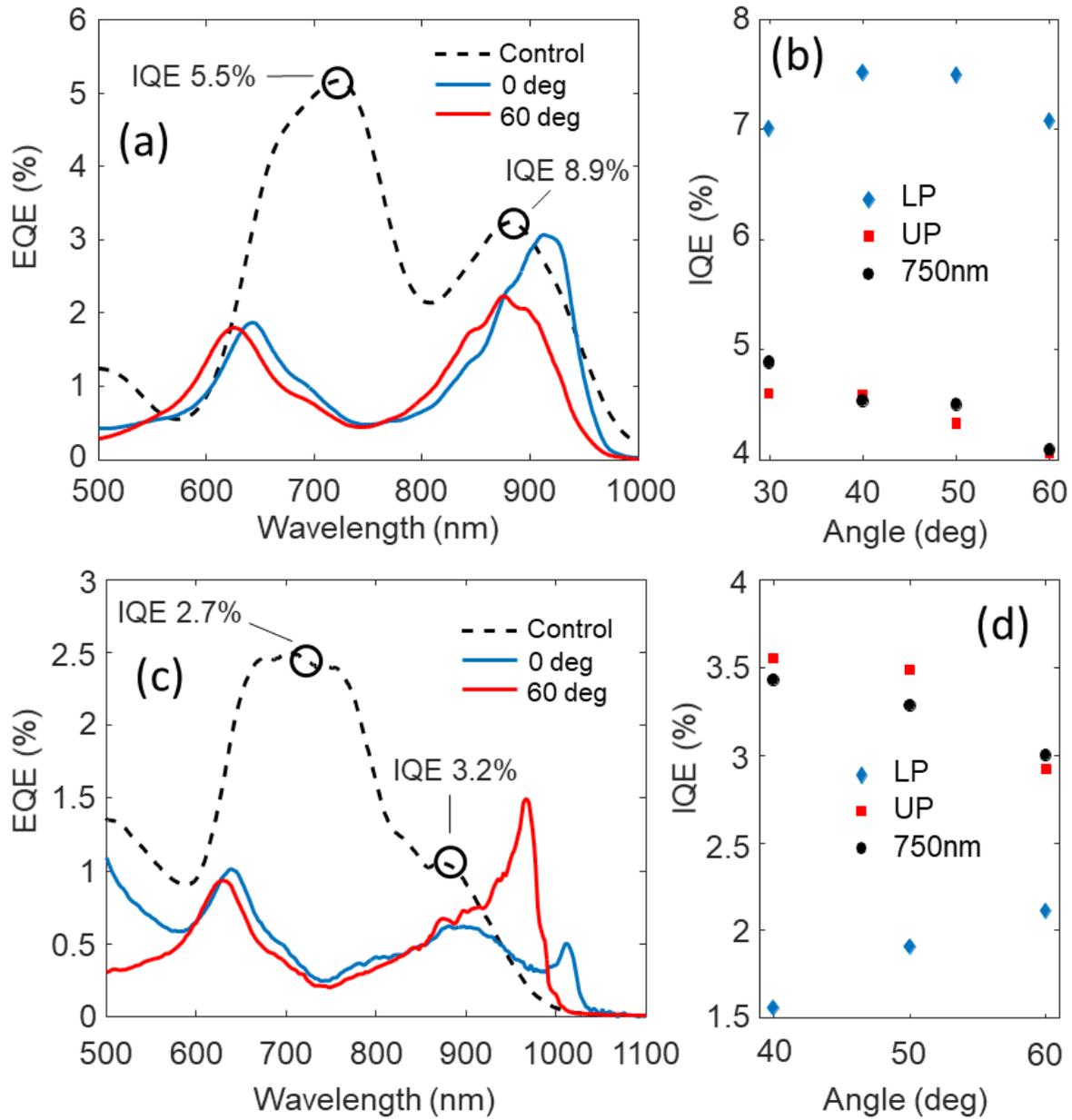

**Figure 5.** (a) EQE spectra and (b) IQE values for the infrared polariton photodiode with 75 nm of mixed layer of SnPc:$C_{60}$. (c) EQE spectra and (d) IQE values for the infrared polariton photodiode with 110 nm of mixed layer of SnPc:$C_{60}$. Blue lines: incident light at normal incident, red lines: at 60 degrees. Dashed black line: EQE spectrum for the control device with normal incident light shined from the ITO-side.

# Supporting Information: Organic Photodiodes with an Extended Responsivity using Ultrastrong Light-Matter Coupling

To study strong coupling of CuPc, we fabricated a microcavity structure consisting of Ag (75 nm), CuPc (150 nm) and Ag (35 nm) on a glass substrate. The angle resolved reflectivity measurements for TE and TM polarizations are shown in Figure S1. For TE polarization, the Rabi splitting found from a least square fit to the Hopfield Hamiltonian is $\hbar\Omega = 0.95 \pm 0.01$ eV with detuning of $\Delta = 0.14 \pm 0.02$ eV. For TM polarization, the found Rabi splitting is $\hbar\Omega = 0.92 \pm 0.01$ eV with detuning of $\Delta = 0.14 \pm 0.01$ eV.

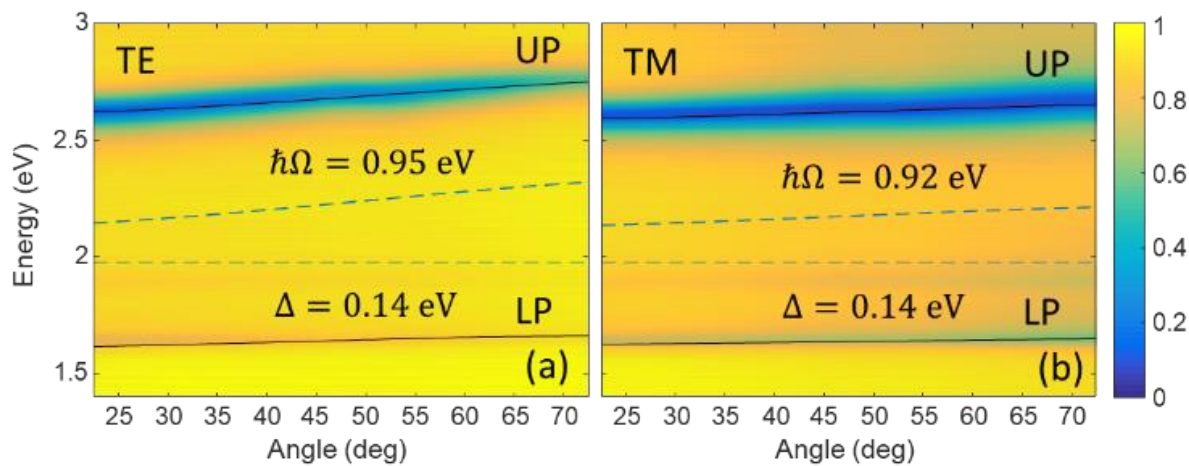

**Figure S1.** Angle resolved reflectivity spectra of CuPc microcavity for (a) TE and (b) TM polarizations. The solid black lines show the least square fit to Hopfield Hamiltonian and the dashed lines are the energies of the corresponding uncoupled photonic and excitonic modes.

We also studied strong coupling of SnPc mixed with $C_{60}$ in a microcavity structure consisting of Ag (75 nm), mixed SnPc:$C_{60}$(160 nm; 70%: 30% by volume) and Ag (20 nm) on a glass substrate. The angle resolved reflectivity results for TE and TM polarizations are shown in Figure S2. For the TE mode we find $\hbar\Omega = 0.78 \pm 0.01$ eV, $\Delta = -0.17 \pm 0.01$ eV and for TM mode $\hbar\Omega = 0.76 \pm 0.01$ eV, $\Delta = -0.19 \pm 0.01$ eV.

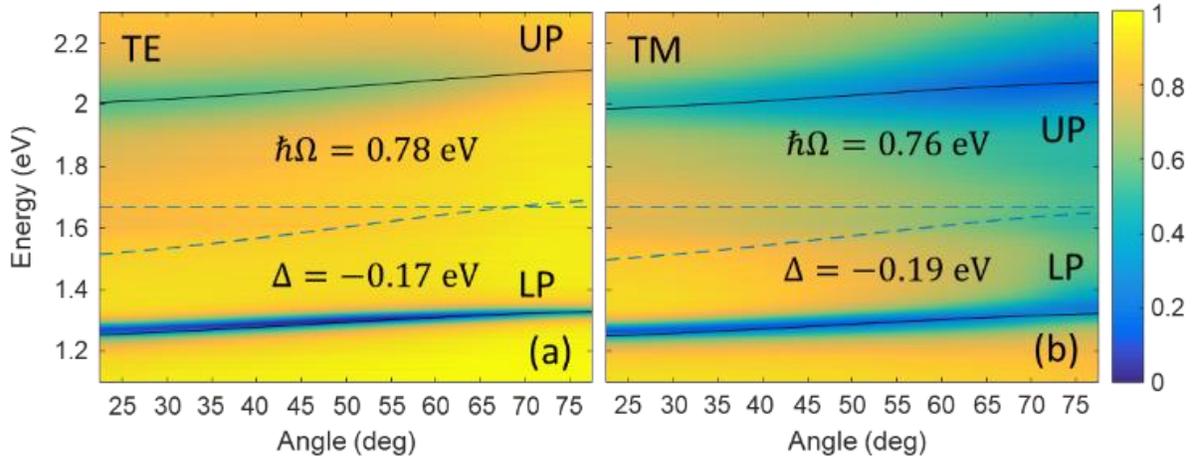

**Figure S2.** Angle resolved reflectivity spectra of mixed SnPc:$C_{60}$ microcavity for (a) TE and (b) TM polarizations. The solid black lines show the least square fit to Hopfield Hamiltonian and the dashed lines are the energies of the corresponding uncoupled photonic and excitonic modes.